\documentclass[twocolumn,showpacs,superscriptaddress, preprintnumbers , amsmath,amssymb,letterpaper,floatfix]{revtex4}
\usepackage{dcolumn}% Align table columns on decimal point
\usepackage{bm}% bold math
\usepackage{amsmath}
\usepackage{multirow}
\usepackage{graphicx}
\usepackage{float}
\usepackage[colorlinks=true]{hyperref}
\begin{document}

 %\twocolumn
 %\baselineskip = 1.5\baselineskip

 \newcommand{\re}{\mathop{\mathrm{Re}}}
 \newcommand{\im}{\mathop{\mathrm{Im}}}
 \newcommand{\D}{\mathop{\mathrm{d}}}
 \newcommand{\I}{\mathop{\mathrm{i}}}
 \newcommand{\E}{\mathop{\mathrm{e}}}
 \newcommand{\unite}[2]{\mbox{$#1\,{\rm #2}$}}
 \newcommand{\myvec}[1]{\mbox{$\overrightarrow{#1}$}}
 \newcommand{\mynor}[1]{\mbox{$\widehat{#1}$}}
 \newcommand{\rmsemit}{\mbox{$\tilde{\varepsilon}$}}
 \newcommand{\mean}[1]{\mbox{$\langle{#1}\rangle$}}

\title{Intense Super-radiant X-rays from a Compact Source using a Nanocathode Array and Emittance Exchange}
% Force line breaks with \\
\author{W. S. Graves}\affiliation{Massachusetts Institute of Technology, Cambridge, MA 02139, USA}
\author{F. X. K\"artner} \affiliation{Massachusetts Institute of Technology, Cambridge, MA 02139, USA}
\affiliation{Center for Free-Electron Laser Science, Deutsches Elektronen-Synchrotron, 22607 Hamburg, Germany}
\author{D. E. Moncton} \affiliation{Massachusetts Institute of Technology, Cambridge, MA 02139, USA}
\author{P. Piot} \affiliation{Department of Physics, Northern Illinois University, DeKalb IL 60115,
USA} \affiliation{Fermi National Accelerator Laboratory, Batavia, IL 60510, USA}
\date{\today}% It is always \today, today,

\begin{abstract}
A novel method of producing intense short wavelength radiation from relativistic electrons is described.  The electrons are periodically bunched at the wavelength of interest enabling in-phase super-radiant emission that is far more intense than from unbunched electrons.  The periodic bunching is achieved in steps beginning with an array of beamlets emitted from a nanoengineered field emission array.  The beamlets are then manipulated and converted to a longitudinal density modulation via a transverse to longitudinal emittance exchange.  Periodic bunching at short wavelength is shown to be possible, and the partially coherent x-ray properties produced by Inverse Compton scattering from an intense laser are estimated.  The proposed method increases the efficiency of x-ray production by several orders of magnitude, potentially enabling compact x-ray sources to produce brilliance and flux similar to major synchrotron facilities.
\end{abstract}
                              %display desired%
\maketitle
Intense x-ray sources including synchrotrons and inverse Compton scattering (ICS) sources rely respectively on the scattering of virtual (static undulator magnetic field) or real photons (laser) into the x-ray spectrum by collision with relativistic electrons.  In these sources electrons are grouped in bunches much longer than the x-ray wavelength with random temporal distribution within the bunch.  The phases of the x-ray wavetrains produced by independent electrons are not correlated, except in the case of the free-electron laser (FEL) in which an instability leads to phase-ordering.  The increase in intensity of FEL radiation over synchrotron radiation is due to in-phase x-ray emission that scales as $N_e^2$ rather than  $N_e$, where $N_e\approx10^7$ is the number of electrons in a bunch.  This Letter describes a method to produce in-phase emission of x-rays without use of the FEL instability.  Instead it begins with an array of beamlets (bunches of tens of electrons) produced by a nanocathode field-emission array (FEA).  The array can be viewed as a periodic {\it transverse} electron density modulation.  The beamlets are accelerated to relativistic energy and transported through an emittance-exchange (EEX) line that converts the transverse density modulation into a {\it longitudinal} modulation that enables in-phase radiation generation.  The properties of the cathode, accelerator and beam transport elements, and super-radiant x-ray emission are each described.

The FEA consists of a two dimensional grid of sharp electron emitters surrounded by gates to regulate emission and focus each of the beamlets.  The sharp tip geometry of each emitter concentrates applied laser or DC fields onto a small area, resulting in field-emitted electrons with ultralow normalized emittance $\varepsilon_{xn}=1/(mc)[\mean{x^2}\mean{p_x^2}-\mean{xp_x}^2]^{1/2}$ where $\mean{x^2}$ and $\mean{p_x^2}$ are the second moments of the beam's transverse position and momentum.  The normalized emittance is a measure of phase space area and is a principal measure of beam quality.  The electric fields in the gate region around a tip are modeled with program POISSON~\cite{poisson} using a tip radius of 3~nm, tip pitch of 500~nm, and gate radius 200 nm.   The applied fields are normal to the local tip surface, peaking at $7\times10^9$~V/m at a gate voltage of 31~V.  Using a Fowler-Nordheim model of emission~\cite{fn,jensen} the peak current density is $9\times10^{11}$ A/m$^2$ with an RMS angular extent of $\sim$ 20~degrees.  Total current/tip is 15~$\mu$A but modeling shows 30-40\% particle loss on the focusing lens gates, yielding an average 10~$\mu$A transmitted current per tip.  Photofield emission~\cite{mustonen} has demonstrated similar current density and will be used to time-gate emission for 1~ps FWHM pulses.  Given the single tip current and time duration, a total charge per pulse of 1.6~pC would be produced by, e.g., a $400\times400$ array.  

\begin{figure}[t]
\includegraphics[width=0.4\textwidth]{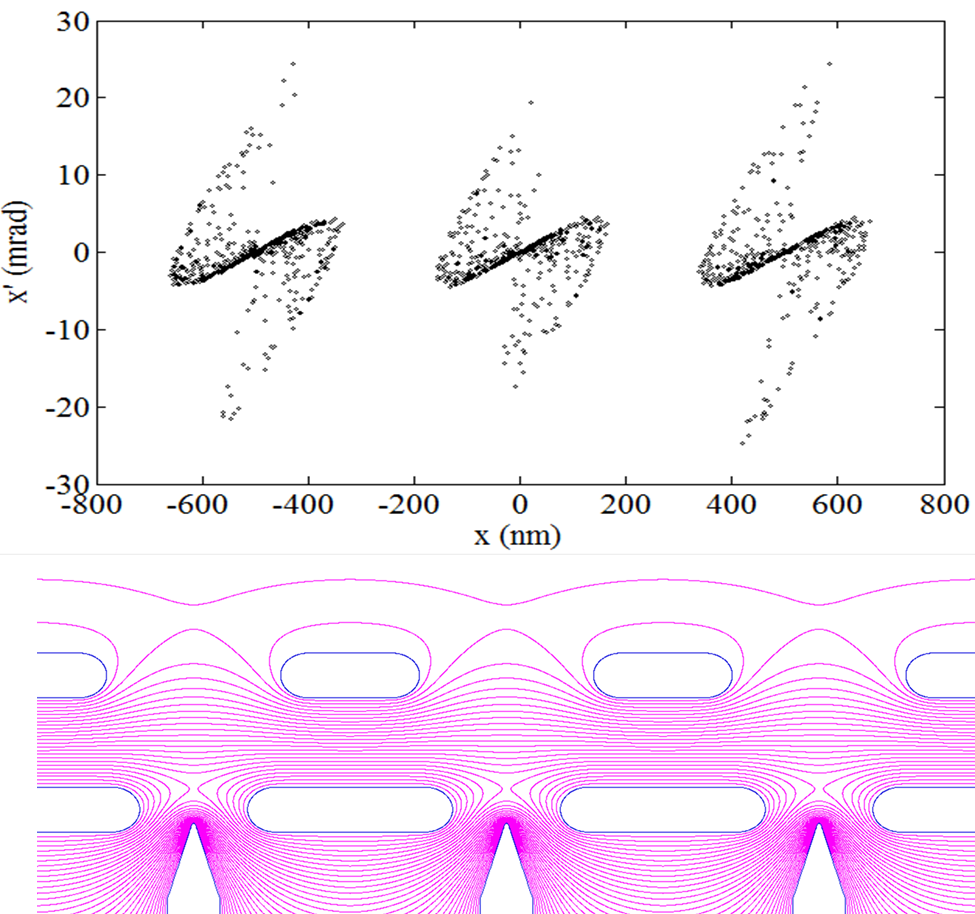}
\caption{Transverse phase space distributions (top) showing dense cores and diffuse tails for 3 beamlets at cathode assembly exit.  Bottom shows POISSON model electrostatic fields in cathode assembly consisting of nanotips and focusing gates.  Fields are cylindrically symmetric around tips.}
\label{fig:tipsbeamlets}
\end{figure}

The RMS source size of each tip is $\sqrt{\mean{x^2}}~=~0.7$~nm.  The uncertainty principle requires $\sqrt{\mean{x^2}\mean{p_x^2}}~\ge~\hbar/2$ yielding minimum transverse momentum spread $\sqrt{\mean{p_x^2}}\ge 140$~eV/c.  This is consistent with measured values~\cite{shimawaki} of FEA electron energy spread from ungated tips.  The low initial emittance is consistent with theory~\cite{dowell, jensen2, floettmann} and measurement~\cite{graves2}.  Aberrations in the electrostatic lenses~\cite{rose} cause emittance growth of 1-2 orders of magnitude so that the emittance at cathode exit is not sensitive to the initial value.  Particle tracking of 100 random ensembles finds that the final single-tip emittance varies from $8 \times 10^{-12}$ to $20\times10^{-12}$~m-rad at the cathode assembly exit.  This emittance growth is tolerable for eventual bunching at soft x-ray wavelengths.  For hard x-rays it will be important to limit the emittance growth to be able to bunch the beam at sub-angstrom lengths.  This effect is described in the beam dynamics section below.  The cathode assembly includes an array of electrostatic Einzel lenses and post acceleration to $\sim 80$~eV over a 3 micron distance.  Each beamlet exits an individual lens, emerging into the RF injector accelerating fields. The charged particle tracking program PARMELA~\cite{parmela} is used first to track individual distributions from each tip through the electrostatic fields of the cathode assembly, and then the entire array of beamlets through the RF accelerator and transport line.  Fig.~\ref{fig:tipsbeamlets} shows the transverse phase space for 3 beamlets after the lenses, each having a dense core with long tails that account for the emittance growth.   The present study tracks a $9\times9$ array of beamlets from emission through transformation by the EEX line.

\begin{figure*}[t]
\includegraphics[width=0.90\textwidth]{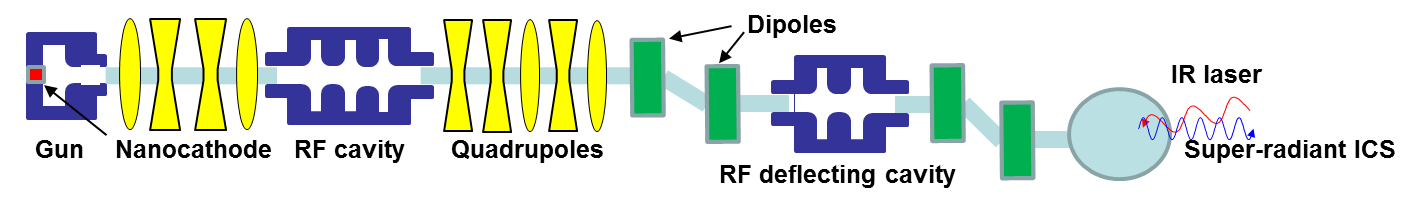}
\caption{Layout of the accelerator components and emittance-exchange beamline.  Total length is 2.3 m.}
\label{fig:layout}
\end{figure*}

The layout of the accelerator components is shown in Fig.~\ref{fig:layout} and consists of a RF gun with nanocathode assembly followed by four quadrupole magnets, then a 3-cell RF cavity to tune the final energy and energy chirp, followed by 5 quadrupoles to complete the EEX matching requirements, and finally the EEX line.  Models of the RF fields in each of the cavities are produced by SUPERFISH using the geometry of a high repetition rate room-temperature structure~\cite{tantawi} operating at 9.3 GHz.  The gun is a half-cell cavity 1.4 cm in length and the 3-cell linac cavity has 6.4 cm length.  The total beamline length including accelerator, magnets, and EEX line is a compact 2.3 m.

\begin{figure}[hb]
\includegraphics[width=0.4\textwidth]{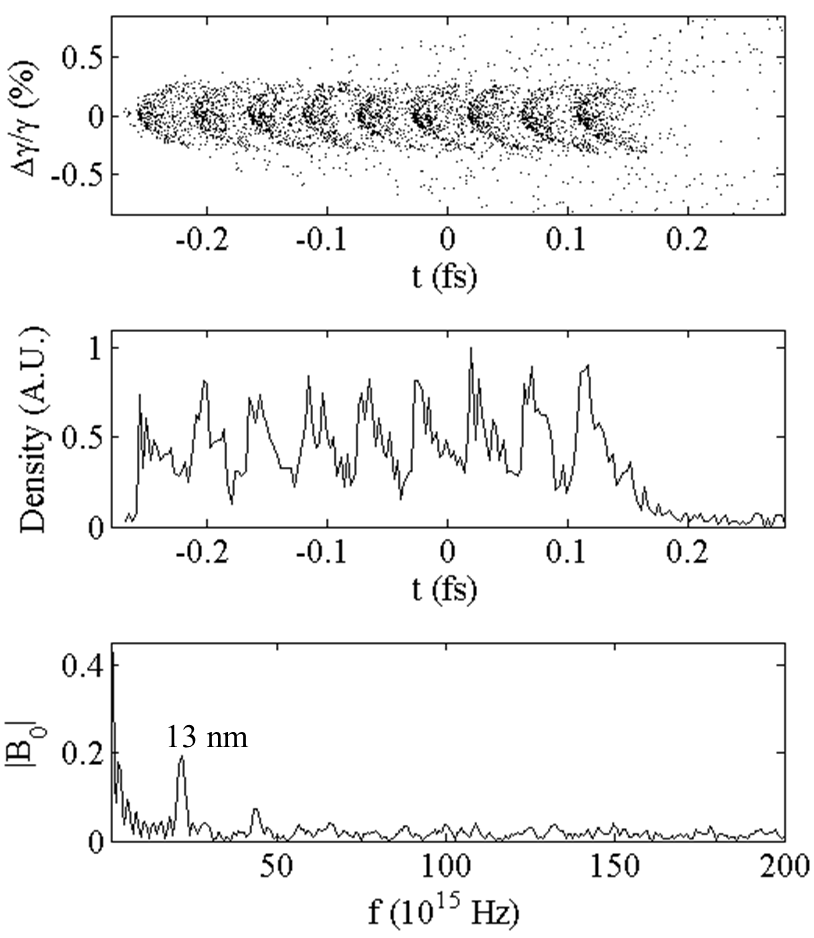}
\caption{Plots of longitudinal phase space (top) of $9\times9$ array of beamlets after EEX, normalized current (middle) showing modulation at 13 nm, and bunching factor (bottom) with peak $B_0 = 0.2$ at $\lambda_x = 13$ nm.}
\label{fig:bunching2}
\end{figure}

The EEX line converts the density modulation from transverse to longitudinal~\cite{cornacchia, emma,carlsten} and has been experimentally demonstrated~\cite{sun} at longer scale lengths.  The line consists of a horizontally-deflecting cavity flanked by two dispersive sections, each with dispersion function~$\eta$. When the deflecting cavity strength is set to $-1/\eta$, the transfer matrix of the beamline in the four-dimensional $(x,x',z,\Delta \gamma/\gamma)$ trace space is block anti-diagonal resulting in an exchange of phase space coordinates between planes.  An undesirable coupling between time and z-position is introduced by the deflecting cavity operating on a TM110 mode, so an accelerating TM010 cell is also incorporated~\cite{zholents} in the structure to cancel the coupling. Tracking and matching of the electron beam through the EEX is performed with code ELEGANT~\cite{elegant} using the PARMELA output. Figure~\ref{fig:bunching2} shows the results of second order tracking of the 9X9 array of beamlets producing clear evidence of modulation.  Simulations excluding second order effects show much deeper modulation, and further optimization will likely improve the results.

The transverse phase space ellipses at the EEX entrance must be matched to produce upright longitudinal ellipses downstream of the EEX that are short relative to the x-ray wavelength $\lambda_x$ with an energy spread low enough that off-energy electrons do not debunch during the laser ICS interaction.  The beamlet length criterion is $\sigma_z \le \lambda_x/4$ and similarly the pathlength tolerance is $\Delta s \le \lambda/4$.  From Eq.~\ref{eq:reson} below, the additional pathlength traveled by an off-energy electron through $N_L$  laser periods is $\Delta s = -2 \lambda_x N_L \Delta \gamma/\gamma$, the negative sign indicating higher energy electrons travel shorter paths. Combining this expression with the pathlength tolerance sets the energy spread tolerance at $\delta \gamma/\gamma \le 1/8 N_L$.  The beamlet length and energy spread limits can be combined to show the longitudinal normalized emittance must satisfy $\varepsilon_{zn} \le \gamma\beta \lambda_x/32 N_L$.  Since $\varepsilon_{zn}$  is the EEX-transformed transverse emittance of a single nanotip, the same limit pertains to the transverse emittance at the nanotip.

For small bend angle $\phi$, the Courant-Snyder parameters $\alpha_x, \beta_x$ must satisfy $\varepsilon_{xn}\beta_x = \gamma\beta(\lambda_x/4\phi)^2$, and $\varepsilon_{xn}\alpha_x = \gamma\beta (\lambda_x/4\phi)^2 /L_{EEX}$where $L_{EEX}$ is the EEX line length.  The EEX line not only transforms, it also compresses the beamlets and their spacing, helping to reach short wavelength, with demagnification $M=\phi < 1$ for small bend angle.  These matching conditions refer to a single beamlet.  There are additional requirements for the ensemble of beamlets that are beyond the scope of this report.  A detailed description of the matching conditions and other electron beam transport issues will be presented in a forthcoming publication\cite{piot}.

The x-ray beam properties are estimated for the intense central cone of radiation in the backscattering configuration.  The resonant wavelength is
\begin{eqnarray}\label{eq:reson}
\lambda_x = \frac{\lambda_L}{4\gamma^2} (1+a_0^2 +\gamma^2 \Delta \theta^2), 
\end{eqnarray}
where $\lambda_x$ is the x-ray walvelength, $\lambda_L$ is the laser wavelength,  $a_0 \ll 1 $ is the dimensionless vector potential, and $\Delta\theta$  is the angle of observation with respect to the electron direction.  ICS radiation shares the physics and properties of undulator radiation in the optimized case where the lengths of the electron bunch and laser pulse are short relative to the Rayleigh length, the electron beam size is smaller than the laser waist, and laser pulse shaping is used to create a flattop time profile. For this case an expression for the intensity of the central cone for a single electron colliding with a linearly polarized laser field is~\cite{krinsky,kim}
\begin{eqnarray}
\left( \frac{d^2U}{d\omega d\Omega} \right)_e\simeq \alpha h a_0^2 N_L^2 \gamma^2, 
\label{eq:du}
\end{eqnarray}
where $\alpha$ is the fine structure constant and $N_L$ is the number of laser periods. The single-electron RMS on-axis bandwidth is  $\Delta\omega/\omega =1/N_L$.  Combining this bandwidth with Eq.~\ref{eq:reson}, the divergence of the central cone is $\Delta\theta = 1/(\gamma \sqrt{N_L})$. The fields generated by individual electrons differ only by a phase factor depending on the initial time  $t_k$ that each electron enters the laser field.  The intensity for many electrons is then given by
\begin{eqnarray}
\left( \frac{d^2U}{d\omega d\Omega} \right)_{tot}  = \left( \frac{d^2U}{d\omega d\Omega} \right)_e \left[N_e + N_e(N_e-1) |B_0|^2 \right]
\label{eq:dutot}
\end{eqnarray}
where the first term in square brackets is the emission from randomly phased electrons and the second term describes super-radiant emission of in-phase electrons.  When the bunching factor $B_0 = \sum\limits_{k=1}^{N_e} e^{i \omega t_k}$ satisfies $B_0 \gg \sqrt{1/N_e}$ then super-radiant emission dominates~\cite{saldin} and the intensity becomes
\begin{eqnarray}
\left( \frac{d^2U}{d\omega d\Omega} \right)_{tot}  \approx  \alpha h a_0^2 N_L^2 \gamma^2 {N_e}^2 |B_0|^2.
\label{eq:dusr}
\end{eqnarray}

The EEX transformation results in a train of  $N_b$ microbunches emitting in phase.  The train narrows the bandwidth of the central cone to $\Delta\omega/\omega = 1/(N_L+N_b)$ and its emission angle to $\Delta\theta=1/(\gamma\sqrt{N_L+N_b})$. Unlike FEL radiation there is no dominant transverse mode so that the output is not generally diffraction limited~\cite{saldin}.  Using this spectrum and angle in Eq.~\ref{eq:dusr}, the total super-radiant energy is
\begin{eqnarray}
U_{tot} &\simeq& \left( \frac{d^2U}{d\omega d\Omega} \right)_{tot} N_e^2 |B_0|^2 \Delta\omega \Delta\Omega \nonumber \\
& \simeq &\pi \alpha E_x a_0^2 \left(\frac{N_L}{N_L+N_b}\right)^2 N_e^2 |B_0|^2, 
\end{eqnarray}
where $\Delta\Omega = \pi\Delta\theta^2$ and photon energy $E_x~=~\hbar \omega$.

From Eq.~\ref{eq:reson} the energy spread required to attain this bandwidth is $\Delta\gamma/\gamma < \Delta \omega/2 \omega$ and the distribution of electron angles due to emittance must satisfy $\Delta\theta_e < (1/\gamma)\sqrt{\Delta\omega/\omega}$, or in terms of emittance $\varepsilon_{xn}~<~\sigma_{ex}/\sqrt{N_L+N_b}$ where $\sigma_{ex}$ is the RMS electron beam size.  

Table~\ref{tab:param} presents the x-ray beam performance of the 13 nm example sources using both the modeled 10 kHz room-temperature gun and a high repetition rate version with cavity-enhanced laser and superconducting RF gun~\cite{harris}.  Such a high repetition rate source would produce about 2 watts of 13 nm power in a high quality, narrow bandwidth optical beam that is well suited to EUV lithography.  Assumed electron and laser parameters are  $N_e=10^7$, $N_b=400$  microbunches, $B_0 = 0.2$, $\gamma\beta = 3.8$ (1.5 MeV), $N_L=190$,  $a_0=0.3$, laser waist size $W_0=6~\mu$m, and laser pulse energy 50~mJ.  The limit on electron beam energy spread is 1.5~keV, and the limit on electron emittance is $\varepsilon_{xn} < 1.2 \times 10^{-7}$ m-rad to meet the bandwidth criteria.
 
\begin{table}[t]
\caption{Estimated performance for 13 nm ICS source. \label{tab:param} }
\begin{center}
\begin{tabular}{l c c c c}\hline\hline\
parameter & Cu linac & SRF linac  &  unit \\
\hline
Photon energy & 92  & 92  & eV  \\
Photons/pulse  &  $8.5\times 10^8$    &  $8.5\times 10^8$       & --  \\
Avg flux (.2\% BW) &  $8.5\times 10^{12}$    &   $1.5\times 10^{17}$       & phot/s  \\
Average power   &  $1.3\times 10^{-4}$    &  2.3  & W  \\
Bandwidth (FWHM) & 0.2 & 0.2 &  \%  \\
Average brilliance  &  $2\times10^{13}$    & $3.4\times10^{17}$   &$^\ast$ \\
Peak brilliance & $2\times10^{23}$  &$2\times10^{23}$ & $^\ast$ \\
Coherent fraction  &  3.5 & 3.5   & \%  \\
Opening angle & 10 &   10 & mrad\\
Source size & 3  & 3 & $\mu$m\\
Pulse length & 26 & 26 & fs  \\
Repetition rate & 10 & 176,000 & kHz \\
Avg electron current & 0.02 & 280 & $\mu$A \\
\hline \hline
\end{tabular}
\end{center}
\vspace{-3mm}
$^\ast$brillance units are phot/(sec 0.1\% mm$^2$ mrad$^2$) 
\end{table}
The performance of a hard x-ray source can be estimated as well assuming more aggressive beam parameters.  For the same laser parameters as above and electron parameters of $N_e = 5\times 10^7$, $N_b = 1000$ microbunches, and $B_0=0.1$, the photon flux at 1 angstrom would be $1.3\times 10^9$ photons per shot into bandwidth of 0.09\%, opening angle of 0.6 mrad and pulse length of 400 attoseconds for peak and average brilliance of $1 \times 10^{28}$ and $1.7 \times 10^{20}$ photons/(sec 0.1\% mm$^2$ mrad$^2$) respectively at a repetition rate of 100 MHz and electron beam energy of 23 MeV. The limit on electron beam energy spread is 10 keV, and the limit on electron emittance to meet the bandwidth criteria is $\varepsilon_{xn} < 9\times 10^{-8}$ m-rad.  Such a source would achieve state-of-the-art performance approaching that of major facilities.

\begin{figure}[ht!]
\includegraphics[width=0.4\textwidth]{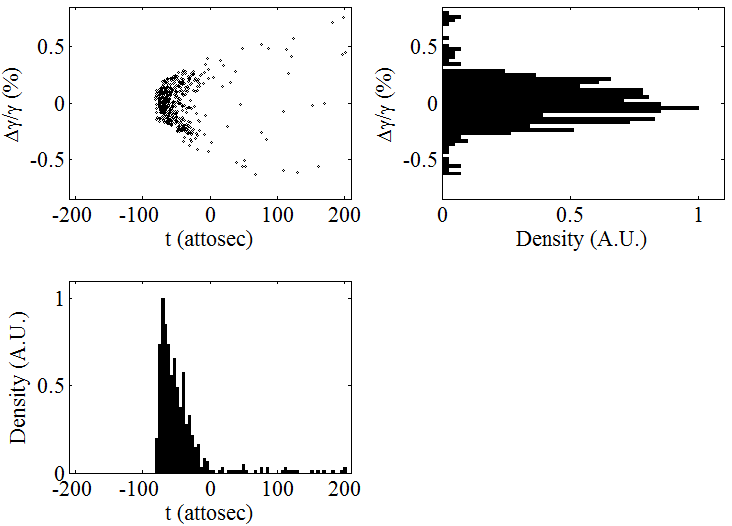}
\caption{Simulation of single period ultrashort bunch created from $1\times9$ nanotip array and EEX transformation. FWHM length is less than 100 attoseconds with RMS energy spread of .09\%.  Use directly for electron microscopy, or for sub-fs x-ray pulse creation.}
\label{fig:onebunch}
\end{figure}

In addition to intense x-ray beams, the combination of nanoarray emission and emittance exchange can produce beams of electrons and x-rays with other interesting properties.  For example the number of periods of bunched beam are controlled by the array dimensions on the nanocathode.  A $1 \times N$ array will produce a beam with N beamlets that emittance-exchange into a single ultrashort microbunch (Fig.~\ref{fig:onebunch}) with low energy spread.  If performed at $\approx$~100 keV the electrons may be used directly for ultrafast microscopy.  Alternatively, such a short pulse can produce an attosecond ICS photon pulse that is $N_L$ x-ray periods long.

In summary, a method has been presented to increase the efficiency, flux and brilliance of a compact x-ray source by orders of magnitude, providing performance that can rival that of major facilities at a small fraction of their cost and size.

We are grateful to Karl Berggren and Luis Velasquez-Garcia for discussions of the field emission array properties.  This work was supported by NSF grant DMR-1042342, DOE grants DE-FG02-10ER46745 and DE-FG02-08ER41532, and DARPA grant N66001-11-1-4192.

\end{document}